# Anomalous Compressibility and Electronic Robustness of Metallic Delafossite $PdCoO_2$ under Pressure

Cheng Peng[1], Pahuni Jain[2], Tyler Harper[2], Chris Leighton[2], Weiwei Xie[1*]

1. Department of Chemistry, Michigan State University, East Lansing, MI, 48824 USA
2. Department of Chemical Engineering and Materials Science, University of Minnesota, Minneapolis, MN, 55455 USA

[*] Corresponding Author: Weiwei Xie (xieweiwe@msu.edu)

## *ABSTRACT*

The layered delafossite $PdCoO_2$ is an exceptional oxide metal whose ultrahigh conductivity arises from a Pd-derived nearly-free-electron band. Here, high-pressure single-crystal X-ray diffraction combined with first-principles calculations is used to investigate its structural, bonding, and electronic evolution up to 10 GPa. $PdCoO_2$ retains the rhombohedral $R$-3$m$ structure throughout the investigated pressure range, with no structural phase transition. The lattice exhibits an unusual anisotropic compression, with the in-plane $a$-axis contracting more strongly than the stacking $c$ axis, opposite to the behavior of most layered materials. Despite this anomalous compressibility, only minor changes are observed in the local Pd-O and Co-O coordination environments, indicating a remarkably rigid bonding framework. Crystal orbital Hamilton population analysis reveals only subtle strengthening of the existing bonding interactions, without the emergence of destabilizing Pd-related antibonding states. Consistent with these findings, the Pd-derived metallic band and quasi-two-dimensional Fermi surface remain essentially unchanged under compression. Boltzmann transport calculations further show that the in-plane conductivity is nearly pressure-independent, whereas the out-of-plane conductivity decreases modestly, resulting in a slight increase in transport anisotropy. These results demonstrate that the anomalous compressibility of $PdCoO_2$ originates from its robust chemical bonding network, which preserves both the crystal structure and the highly conductive Pd-derived metallic state under pressure.

## I. INTRODUCTION

Layered materials generally exhibit highly anisotropic mechanical responses because their constituent layers are connected through relatively weak interlayer interactions. As a result, compression is usually accommodated predominantly along the crystallographic *c*-axis, while the in-plane framework remains comparatively rigid. This characteristic behavior is widely observed in van der Waals materials, transition-metal dichalcogenides, graphite, and many layered oxides[1–3], where the large difference between intra- and inter-layer bonding governs both the structural evolution and pressure-induced electronic phenomena. Identifying layered compounds that deviate from this conventional compression mechanism is therefore important for understanding how chemical bonding dictates mechanical stability and electronic robustness under extreme conditions.

Delafossite oxides, with the general formula $ABO_2$, provide an ideal platform for exploring the above issues[4–6]. Their crystal structures consist of triangular layers of linearly coordinated monovalent A-site cations alternating with edge-sharing $BO_6$ octahedral layers, producing a layered architecture that hosts a remarkable diversity of electronic ground states. While many delafossites are insulating or semiconducting, including the frustrated magnet $CuFeO_2$[7] and the transparent conducting oxide $CuAlO_2$[8], a small family of Pd- and Pt-based compounds, including $PdCoO_2$, $PtCoO_2$, and $PdCrO_2$, are among the most conductive oxide metals known[9]. In these materials, electronic transport is dominated by nearly-free-electron states derived from the noble-metal layer, whereas the transition-metal oxide slabs remain comparatively ionic and electronically inactive[10,11]. Among this family, $PdCoO_2$ has emerged as a model system for studying low-dimensional metallic transport[12–15]. Its electronic structure is characterized by a single Pd-derived band crossing the Fermi level, which produces an almost perfectly hexagonal Fermi surface and extraordinarily long electronic mean free paths[16]. The conventional ionic picture describes Pd as monovalent $Pd^{+}$ and Co as low-spin $Co^{3+}$ with a filled $t_{2g}$ manifold, resulting in highly conductive Pd layers separated by nonmagnetic $CoO_6$ slabs[17,18]. Although $PdCoO_2$ possesses a crystallographically layered structure, the nearly linear O-Pd-O coordination linking adjacent layers suggests that its interlayer bonding may differ fundamentally from that of conventional layered compounds. Whether this unusual bonding network produces an unconventional mechanical response under compression remains open.

External pressure provides a powerful means of probing such issues because it continuously modifies interatomic distances, orbital overlap, and chemical bonding, without altering chemical composition[19]. If $PdCoO_2$ were to behave like a typical layered material, one would expect the weakly bonded stacking direction to accommodate most of the lattice contractions. However, the rigid O-Pd-O dumbbells and highly connected $CoO_2$ framework may instead redistribute the applied stress, generating anomalous compressibility and challenging the conventional understanding of layered oxides. Such behavior would have important implications for the stability of the Pd-derived metallic band and for understanding how chemical bonding controls electronic transport under pressure. Previous high-pressure powder X-ray diffraction measurements demonstrated that $PdCoO_2$ retains the rhombohedral $R$-3$m$ delafossite structure below 10 GPa and exhibits anisotropic lattice compression[20]. However, the origin of this unusual compressibility remains unresolved. Powder diffraction primarily determines the average evolution of the unit cell and provides limited information about pressure-induced changes in atomic coordinates, local coordination environments, and bond-specific structural responses. Consequently, the relationship between the anomalous lattice compression, the rigidity of the Pd conducting layer and $CoO_2$ framework, and the persistence of the Pd-derived electronic states remains to be established.

Here, we combine high-pressure single-crystal X-ray diffraction with first-principles calculations to investigate the structural, bonding, and electronic evolution of single-crystalline $PdCoO_2$ up to 10 GPa. We show that, despite its layered crystal structure, $PdCoO_2$ exhibits an unconventional compression mechanism in which the in-plane $a$-axis contracts more strongly than the stacking $c$-axis, opposite to the behavior of most layered materials. Bond-length analysis reveals that both the linear O-Pd-O units and the $CoO_2$ framework remain remarkably rigid under compression. Density functional theory, crystal orbital Hamilton population analysis, electron localization function calculations, and Boltzmann transport simulations further demonstrate that this anomalous compressibility is accompanied by minimal changes in chemical bonding, and preservation of the characteristic Pd-derived metallic electronic structure. These results establish that $PdCoO_2$ is not simply a structurally stable layered metal but a chemically robust, pressure-resilient delafossite whose unusual compression mechanism originates from its strong three-dimensional bonding network despite its layered crystal architecture.

## II. EXPERIMENTAL SECTION

**A. Crystal synthesis.** $PdCoO_2$ single crystals were prepared following the two-step metathesis and chemical vapor transport route previously reported for high-quality $PdCoO_2$ crystals[15,18]. First, $PdCoO_2$ precursor powder was prepared from high-purity $PdCl_2$, $Co_3O_4$, and Pd powders using the metathesis reaction $PdCl_2 + Co_3O_4 + Pd \rightarrow 2PdCoO_2 + CoCl_2$. The starting powders were mixed in a 1.5:1:1 molar ratio, sealed under vacuum in quartz ampoules, heated to 750 °C, held for 150 h, cooled to 400 °C, and then furnace-cooled to room temperature. Residual chlorides were removed by washing the product in boiling ethanol and acetone. Second, $PdCoO_2$ crystals were grown by $PdCl_2$-assisted chemical vapor transport. Approximately 1 g of precursor was sealed with $PdCl_2$ in an evacuated quartz ampoule and placed in a two-zone furnace with the hot and cold zones held at 760 and 710 °C. After an initial inversion period, the growth continued for 13 d to obtain $PdCoO_2$ crystals.

**B. Ambient pressure single-crystal X-ray diffraction (SCXRD).** A selected $PdCoO_2$ crystal ($0.07 \times 0.06 \times 0.06$ mm$^3$) was mounted on a nylon loop with Paratone oil and measured at ambient conditions using a Rigaku XtaLAB Synergy Dualflex diffractometer equipped with a HyPix detector and a microfocus Mo Kα radiation source (λ = 0.71073 Å, 50 kV, 1 mA). Data were collected using ω scans, with optimal strategies generated by CrysAlisPro (Rigaku OD, version 1.171.42.101a, 2023). Data reduction included Lorentz and polarization corrections. Absorption corrections were applied using numerical Gaussian integration over a multifaceted crystal model[21], followed by empirical spherical harmonics scaling (SCALE3 ABSPACK)[22]. The structure was solved and refined using the SHELXT/SHELXL software package[23,24].

**C. High-pressure SCXRD.** High-pressure measurements were performed on the same $PdCoO_2$ crystal characterized at ambient pressure, at room temperature, using a Diacell One20DAC (Almax-easyLab) with 500 μm culet Boehler-Almax anvils. The single crystal was loaded into the stainless-steel gasket hole (200 μm in diameter) together with a 4:1 methanol/ethanol pressure-transmitting medium to maintain quasi-hydrostatic conditions. Small ruby chips were loaded near the crystal, and the pressure was determined by the ruby fluorescence method.

**D. Electronic structure calculations.** Density functional theory (DFT) calculations were performed using Quantum ESPRESSO (v7.4.1)[25,26]. The calculations utilized projector augmented-wave (PAW) pseudopotentials in conjunction with the Perdew-Burke-Ernzerhof (PBE) exchange-correlation functional[27,28]. A wavefunction cutoff energy of 100 Ry and a charge density cutoff set to ten times this value were employed. For the pressure-dependent electronic-structure, chemical-bonding, electron-localization, and Boltzmann transport analyses, experimentally determined crystal structures were used. Separate structural relaxations were performed with the same computational settings only to compare calculated and experimental lattice parameters. For the primitive-cell electronic-structure calculations and the hexagonal conventional-cell transport calculations, 18 × 18 × 18 and 18 × 18 × 3 Monkhorst–Pack *k*-point meshes were used for self-consistent-field calculations, respectively[29]. Non-self-consistent-field calculations were then carried out on denser 40 × 40 × 40 and 36 × 36 × 6 *k*-point meshes for the primitive and hexagonal conventional cells, respectively. The Davidson diagonalization algorithm was applied[30], and the convergence threshold for self-consistency was set to $10^{-9}$ Ry. The high-symmetry path for the Brillouin zone was generated using the Spglib library[31,32]. Spin-polarized calculations were performed to test possible magnetic instability. The calculations converged to a nonmagnetic solution with negligible total and local magnetic moments, consistent with the experimentally reported nonmagnetic metallic behavior of $PdCoO_2$. Inclusion of spin-orbit coupling (SOC) was tested for the ambient-pressure structure and is discussed in the Supplemental Material. Because SOC does not qualitatively alter the band dispersion near the Fermi level, scalar-relativistic calculations were used for pressure-dependent analyses. Crystal orbital Hamilton population (COHP) analyses were conducted using LOBSTER (v5.1.1)[33]. The basis functions included Pd 5*s*, 5*p*, and 4*d*; Co 3*s*, 4*s*, 3*p*, 4*p*, and 3*d*; and O 2*s* and 2*p* states. COHP curves were calculated over an energy window from −10 to 5 eV relative to the Fermi level. For each pressure point, equivalent nearest-neighbor interactions were grouped by pair type, and average −ICOHP values were used to compare the pressure dependence of the bonding interactions. Conductivity components were calculated using BoltzTraP2 at 300 K within the constant relaxation time approximation[34]. The hexagonal conventional cell was used for the conductivity-tensor calculations so that $\sigma_{xx}$ and $\sigma_{yy}$ correspond to in-plane components and $\sigma_{zz}$ corresponds to the out-of-plane component. The in-plane conductivity was defined as $\sigma_{\text{in-plane}}/\tau = (\sigma_{xx}/\tau + \sigma_{yy}/\tau)/2$.

# III. RESULTS AND DISCUSSION

## A. Pressure evolution of the crystal structure

Single-crystal X-ray diffraction measurements were performed to investigate the structural evolution of $PdCoO_2$ under compression at ambient pressure and 3.7, 6.5, 7.8, and 9.9 GPa. The crystallographic refinement details and atomic parameters are summarized in **Tables I** and **II**. At all measured pressures, $PdCoO_2$ retains the rhombohedral delafossite structure with space group *R*-3*m*, demonstrating that no pressure-induced crystallographic phase transition occurs below 10 GPa. The refined crystal structure remains fully consistent with the ambient-pressure phase, consisting of triangular Pd layers linked by linear O-Pd-O coordination and separated by edge-sharing $CoO_6$ octahedral layers to form the characteristic layered delafossite framework (**Figure 1**). No additional diffraction reflections, peak splitting, or symmetry lowering are observed upon compression, and the refined atomic positions exhibit only negligible changes throughout the investigated pressure range. These results demonstrate that the pressure response of $PdCoO_2$ is governed by continuous lattice compression while preserving its crystallographic symmetry and local structural framework.

**Table I.** The crystal structure and refinement parameters of $PdCoO_2$ at 300 K and ambient pressure, and under 3.7, 6.5, 7.8 and 9.9 GPa. Values in parentheses are estimated standard deviations from refinement.

| Pressure (GPa) | Ambient | 3.7 | 6.5 | 7.8 | 9.9 |
|---|---|---|---|---|---|
| Space Group | *R-3m* | *R-3m* | *R-3m* | *R-3m* | *R-3m* |
| Unit Cell dimensions (Å) | a = 2.8304(2)<br>c = 17.7683(18) | a = 2.8104(2)<br>c = 17.7037(12) | a = 2.7948(2)<br>c = 17.6530(14) | a = 2.7926(3)<br>c = 17.6440(15) | a = 2.7817(4)<br>c = 17.604(2) |
| Volume (Å$^3$) | 123.27(2) | 121.096(19) | 119.41(2) | 119.16(2) | 117.97(4) |
| Z | 3 | 3 | 3 | 3 | 3 |
| Density (calculated) | 7.974 | 8.118 | 8.232 | 8.250 | 8.333 |
| Absorption coefficient (mm$^{-1}$) | 20.370 | 20.897 | 21.191 | 21.239 | 21.451 |
| F(000) | 267 | 267 | 267 | 267 | 267 |
| 2$\theta$ range | 13.784 to 81.366 | 13.834 to 73.464 | 13.874 to 74.368 | 13.88 to 74.06 | 13.914 to 72.88 |
| Reflections collected | 698 | 1050 | 637 | 339 | 545 |
| Independent reflections | 132 [$R_{int}$ = 0.0658] | 101 [$R_{int}$ = 0.1031] | 102 [$R_{int}$ = 0.1369] | 98 [$R_{int}$ = 0.0343] | 95 [$R_{int}$ = 0.1153] |
| Data/restraints/parameters | 132/0/9 | 101/0/9 | 102/0/9 | 98/0/9 | 95/0/9 |

| Final $R$ indices ($I$>2σ($I$)) | $R_1$ = 0.0309; $wR_2$ = 0.0802 | $R_1$ = 0.0472; $wR_2$ = 0.1108 | $R_1$ = 0.0816; $wR_2$ = 0.1688 | $R_1$ = 0.0449; $wR_2$ = 0.1104 | $R_1$ = 0.0756; $wR_2$ = 0.1657 |
|---|---|---|---|---|---|
| Final $R$ indices (all) | $R_1$ = 0.0309; $wR_2$ = 0.0802 | $R_1$ = 0.0826; $wR_2$ = 0.1862 | $R_1$ = 0.1277; $wR_2$ = 0.2659 | $R_1$ = 0.0598; $wR_2$ = 0.1333 | $R_1$ = 0.1053; $wR_2$ = 0.2573 |
| Largest diff. peak and hole ($e^-$/Å$^3$) | 1.82 / -2.01 | 8.97 / -10.75 | 10.12/-17.78 | 4.52/-4.92 | 6.89 / -14.34 |
| Goodness-of-fit | 1.215 | 1.440 | 1.452 | 0.919 | 1.393 |

**Table II.** Atomic coordinates and equivalent isotropic atomic displacement parameters ($U_{eq}$, in Å$^2$) of $PdCoO_2$ at 300 K and ambient pressure, and under 3.7, 6.5, 7.8 and 9.9 GPa. ($U_{eq}$ is defined as one-third of the trace of the orthogonalized $U_{ij}$ tensor.)

| P (GPa) | Atom | Wyck. | $x$ | $y$ | $z$ | Occ. | $U_{eq}$ (Å$^2$) |
|---|---|---|---|---|---|---|---|
| Ambient | Pd | 3$a$ | 0 | 0 | 0 | 1 | 0.0050(3) |
| | Co | 3$b$ | 0 | 0 | ½ | 1 | 0.0037(3) |
| | O | 6$c$ | 0 | 0 | 0.11213(2) | 1 | 0.0045(6) |
| 3.7 | Pd | 3$a$ | 0 | 0 | 0 | 1 | 0.0063(15) |
| | Co | 3$b$ | 0 | 0 | ½ | 1 | 0.0043(14) |
| | O | 6$c$ | 0 | 0 | 0.1121(1) | 1 | 0.0110(4) |
| 6.5 | Pd | 3$a$ | 0 | 0 | 0 | 1 | 0.010(2) |
| | Co | 3$b$ | 0 | 0 | ½ | 1 | 0.009(2) |
| | O | 6$c$ | 0 | 0 | 0.11133(2) | 1 | 0.0034(8) |
| 7.8 | Pd | 3$a$ | 0 | 0 | 0 | 1 | 0.0008(8) |
| | Co | 3$b$ | 0 | 0 | ½ | 1 | 0.007(3) |
| | O | 6$c$ | 0 | 0 | 0.1115(1) | 1 | 0.0110(4) |
| 9.9 | Pd | 3$a$ | 0 | 0 | 0 | 1 | 0.0085(18) |
| | Co | 3$b$ | 0 | 0 | ½ | 1 | 0.0074(18) |
| | O | 6$c$ | 0 | 0 | 0.1116(2) | 1 | 0.019(7) |

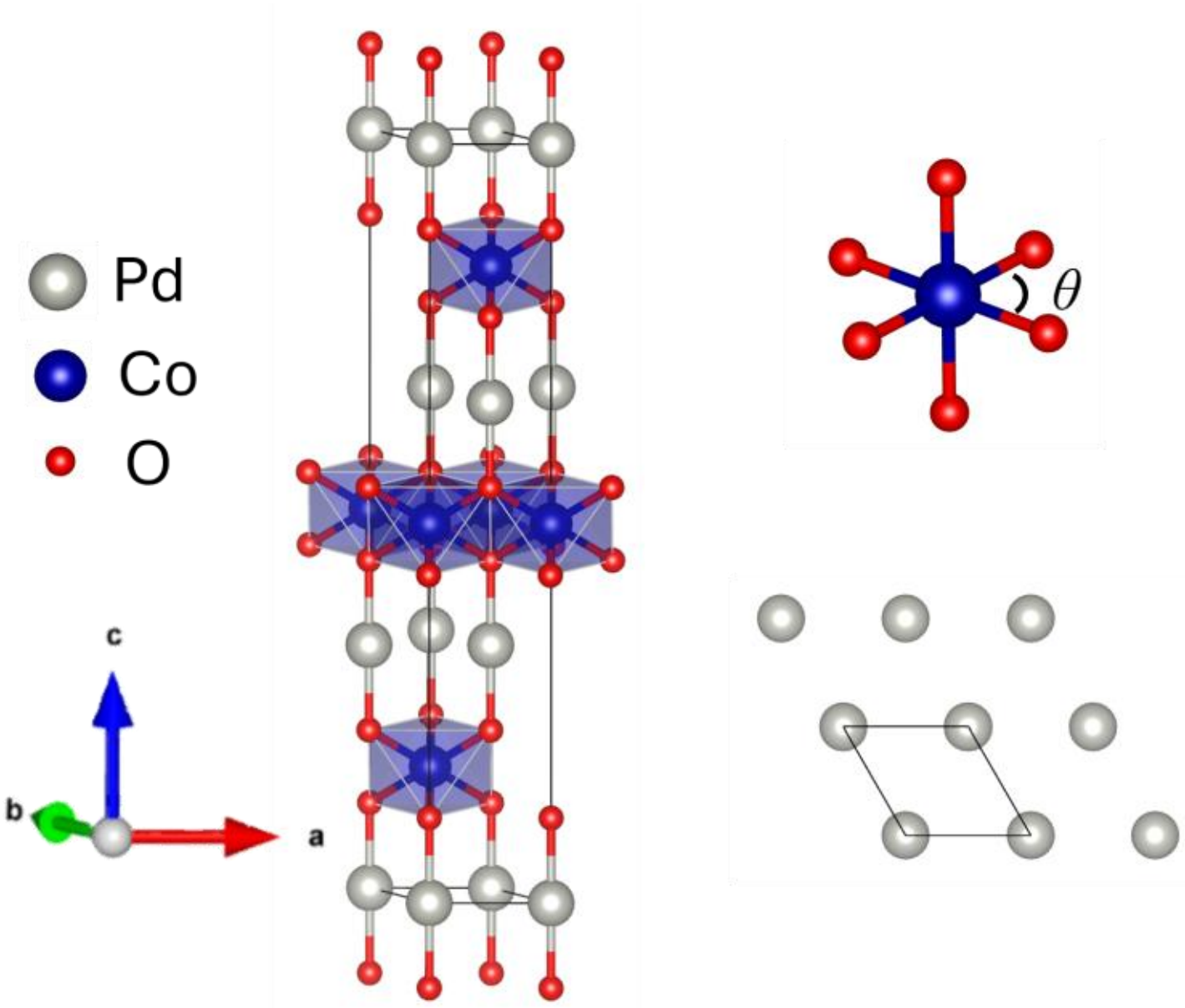

**Figure 1.** Crystal structure of the layered delafossite $PdCoO_2$. **Left:** Perspective view of the rhombohedral $R$-3$m$ crystal structure, showing triangular Pd layers (gray) interconnected by linear O-Pd-

O dumbbells and separated by edge-sharing $CoO_6$ octahedra (blue polyhedra; Co, blue; O, red). **Upper right:** Local octahedral coordination of Co by six oxygen atoms ($CoO_6$). **Lower right:** Top view of the triangular Pd sublattice within the *ab*-plane.

### B. Anomalous anisotropic compressibility under pressure

The refined lattice parameters decrease smoothly with increasing pressure (See Supplemental Material **Figure S1** and **Table SI**). The *a*-lattice parameter decreases from 2.8304(2) Å at ambient pressure to 2.7817(4) Å at 9.9 GPa, corresponding to a contraction of approximately 1.7%. Over the same pressure range, the *c*-lattice parameter decreases from 17.768(2) Å to 17.604(2) Å, corresponding to a smaller contraction of approximately 0.9%. The unit-cell volume decreases monotonically from 123.27(2) $Å^3$ to 117.97(4) $Å^3$. These trends are consistent with the previously reported high-pressure powder diffraction results[20], confirming that the average delafossite framework remains stable under compression. The normalized lattice parameters reveal an anisotropic compression behavior. The in-plane *a*-axis is more compressible than the out-of-plane *c* axis, with approximate linear compressibilities of $k_a = 1.7 \times 10^{-3}$ $GPa^{-1}$ and $k_c = 0.9 \times 10^{-3}$ $GPa^{-1}$. As a result, the *c*/*a* ratio increases from 6.2777 at ambient pressure to 6.3285 at 9.9 GPa (**Figure S1(d)**). This behavior indicates that compression below 10 GPa primarily contracts the triangular in-plane lattice while preserving the overall layered stacking along the *c*-direction.

### C. Equation of state and mechanical rigidity

To quantify the mechanical response of $PdCoO_2$ under compression, the pressure-volume (*P*-*V*) data were fitted using a third-order Birch-Murnaghan equation of state (**Figure 2**),

$$P(V) = 3\,B_0/2\left[(V_0/V)^{7/3} - (V_0 - V)^{5/3}\right]\left\{+(3/4)(B_0{}' - 4)\left[(V_0/V)^{2/3} - 1\right]\right\}.$$

The fitted bulk modulus and its pressure derivative are $B_0$ = 193(23) GPa and $B_0{}'$ = 7(1), respectively. The smooth evolution of the pressure-volume relationship, together with the good agreement between the experimental data and the Birch-Murnaghan fit, provides further evidence that $PdCoO_2$ undergoes continuous elastic compression without any indication of a pressure-induced structural instability up to 10 GPa. The relatively large bulk modulus demonstrates that $PdCoO_2$ is mechanically rigid despite its layered crystal structure. Combined with the anomalous axial compressibility discussed above, these results indicate that the overall incompressibility of $PdCoO_2$ originates from a robust three-dimensional bonding framework rather than the weak interlayer interactions typical of conventional layered compounds. The equation-of-state analysis therefore quantitatively supports the conclusion that the unusual pressure response of $PdCoO_2$ may originate from its distinctive chemical bonding network.

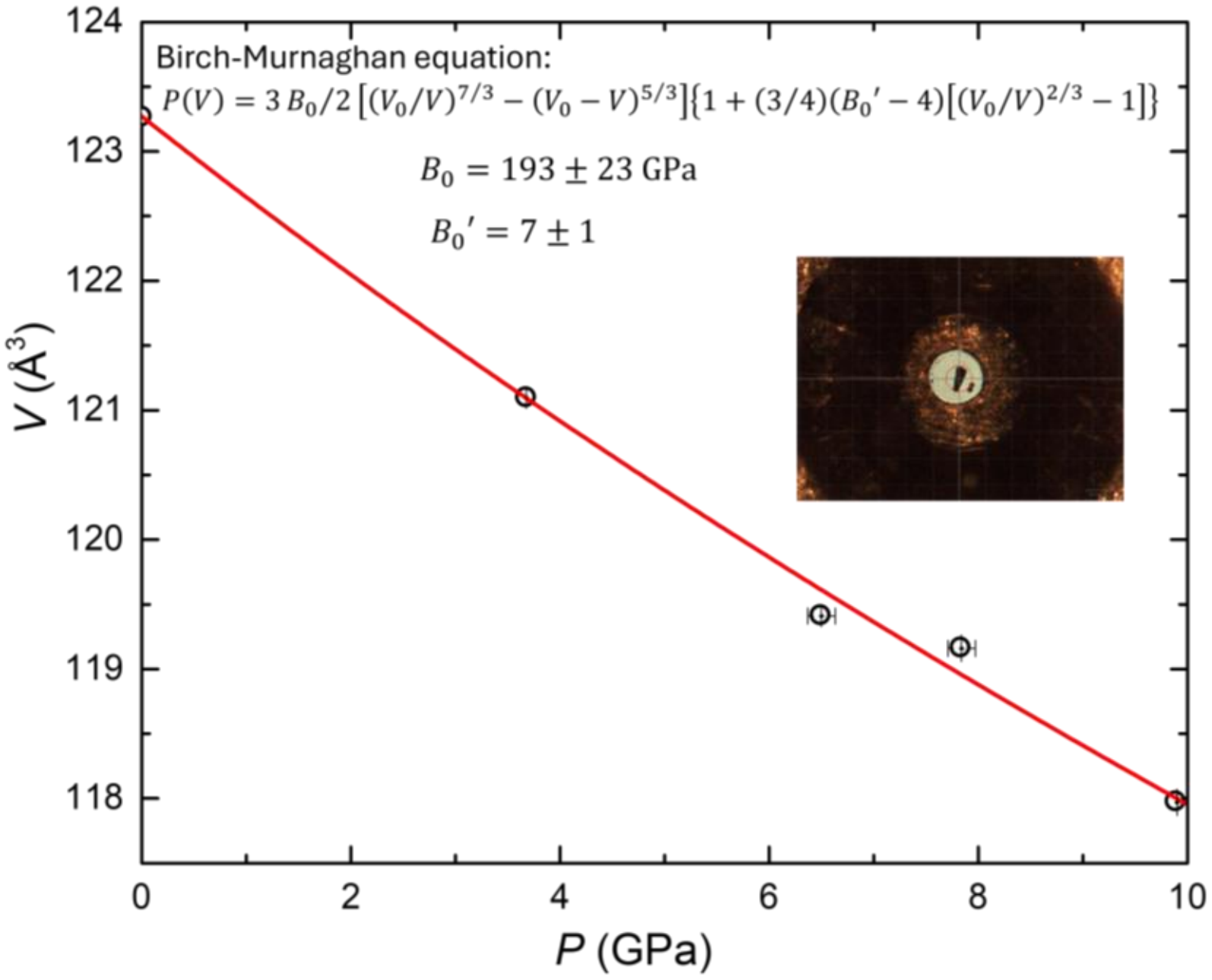

**Figure 2.** Pressure dependence of the unit-cell volume of $PdCoO_2$ measured by high-pressure single-crystal X-ray diffraction. The solid red curve represents the fit to a third-order Birch-Murnaghan equation of state, yielding a bulk modulus of $B_0 = 193(23)$ GPa and a pressure derivative of $\mathbf{B}_0' = 7(1)$. The smooth pressure-volume evolution confirms the absence of a pressure-induced structural phase transition up to 9.9 GPa. **Right inset:** Optical image of the diamond anvil cell showing the sample loaded in the pressure chamber.

### D. Microscopic origin of the anomalous compressibility

High-pressure single-crystal X-ray diffraction enables the evolution of individual interatomic distances and local coordination environments to be determined directly. The pressure dependence of selected bond lengths and bond angles is summarized in the Supplemental Material **Figure S2** and **Table SII**. Overall, all interatomic distances evolve smoothly with increasing pressure, consistent with the continuous lattice contraction discussed above and the absence of a pressure-induced structural transition. The largest bond contractions occur within the basal plane. The Pd-Pd and Co-Co nearest-neighbor distances decrease from 2.8304 Å at ambient pressure to 2.7817 Å at 9.9 GPa, mirroring the evolution of the lattice parameter *a*.

Likewise, the Pd-Co separation decreases nearly linearly from 3.3823 Å to 3.3448 Å. In contrast, the local coordination polyhedra exhibit considerably smaller structural changes. The average Co-O bond length decreases only slightly from 1.900(2) Å to 1.876(2) Å, while the Pd-O bond length remains essentially unchanged within experimental uncertainty throughout the investigated pressure range. Similarly, the Co-O-Co bond angle $\theta$ varies by less than 1°, indicating that the $CoO_6$ octahedra experience little distortion during compression.

These bond-level observations demonstrate that pressure is accommodated primarily through contraction of the triangular Pd lattice rather than significant deformation of the local coordination environments. The nearly invariant Pd-O bond length, together with the minimal changes in the Co-O bonds and Co-O-Co bond angles, indicates that the linear O-Pd-O units and edge-sharing $CoO_6$ octahedra form a remarkably rigid structural framework. This behavior provides a microscopic explanation for the anomalous anisotropic compressibility observed above. Unlike conventional layered materials, where compression is accommodated by shortening weak interlayer contacts, $PdCoO_2$ resists compression along the stacking direction because the strong O-Pd-O linkages remain largely incompressible. Consequently, the lattice accommodates external pressure predominantly through contraction within the *ab*-plane while preserving the overall three-dimensional bonding network and layered crystal architecture.

To further understand the microscopic origin of these effects, crystal orbital Hamilton population (COHP) calculations were performed for the representative nearest-neighbor interactions. The pressure-dependent COHP curves and average integrated COHP (-ICOHP) values are shown in Supplemental Material **Figures S3-S5** and summarized in **Table SIII**. The overall COHP profiles remain remarkably similar throughout the investigated pressure range, indicating that compression does not induce a qualitative reconstruction of the bonding network. In particular, no new antibonding states emerge near the Fermi level for any of the Pd-related interactions, demonstrating that the metallic conduction network remains electronically stable under compression. Although the bonding topology is preserved, the bond strengths evolve selectively with pressure. The Co-O interaction exhibits the largest enhancement, with the average -ICOHP increasing by approximately 9%, indicating appreciable strengthening of the $CoO_6$ framework upon compression. In comparison, the Pd-Pd and Co-Co metallic interactions increase only modestly by approximately 1.7% and 2.4%, respectively, reflecting the shorter in-plane metal–metal separations. The Pd-Co interaction remains essentially unchanged, while the Pd-O interaction changes mainly by less than 1%, consistent with the nearly invariant Pd-O bond lengths obtained from the structural refinements. The excellent agreement between the diffraction and COHP analyses demonstrates that the linear O-Pd-O units remain remarkably rigid even under compression. Taken together, the structural refinements and chemical-bonding analysis reveal that the anomalous compressibility of $PdCoO_2$ originates from its distinctive bonding network. $PdCoO_2$ possesses robust O-Pd-O linkages that effectively resist contraction along the stacking direction. Compression is therefore accommodated primarily by shortening the in-plane Pd triangular lattice, and a modest strengthening of the Co-O framework, while preserving both the local coordination environments and the Pd-derived metallic bonding. The absence of pressure-induced antibonding states or significant bond rearrangements further demonstrates that the anomalous lattice response is governed by the intrinsic rigidity of the chemical bonding network rather than by an electronic instability.

**E. Electron localization function analysis**

To further visualize the real-space electronic distribution, electron localization function (ELF) calculations were performed for $PdCoO_2$ (**Figure 3**). The ELF maps reveal two distinctly different bonding characteristics within the crystal structure. The $CoO_2$ framework exhibits pronounced electron localization around the oxygen atoms, indicative of strong and directional Co-O bonding. In contrast, the Pd layer displays a much more delocalized electron distribution, consistent with metallic bonding within the triangular Pd network and the highly dispersive Pd-derived conduction band. The coexistence of localized Co-O bonding and delocalized Pd electronic states provides a real-space picture of the dual bonding character of $PdCoO_2$. The localized Co-O interactions form a mechanically robust framework that resists structural distortion, whereas the delocalized Pd electrons preserve the metallic conduction network. Together with the single-crystal diffraction and COHP analyses, the results demonstrate that pressure strengthens the existing bonding network without altering its fundamental character, providing a microscopic explanation for the anomalous anisotropic compressibility and the remarkable electronic robustness of $PdCoO_2$ under compression.

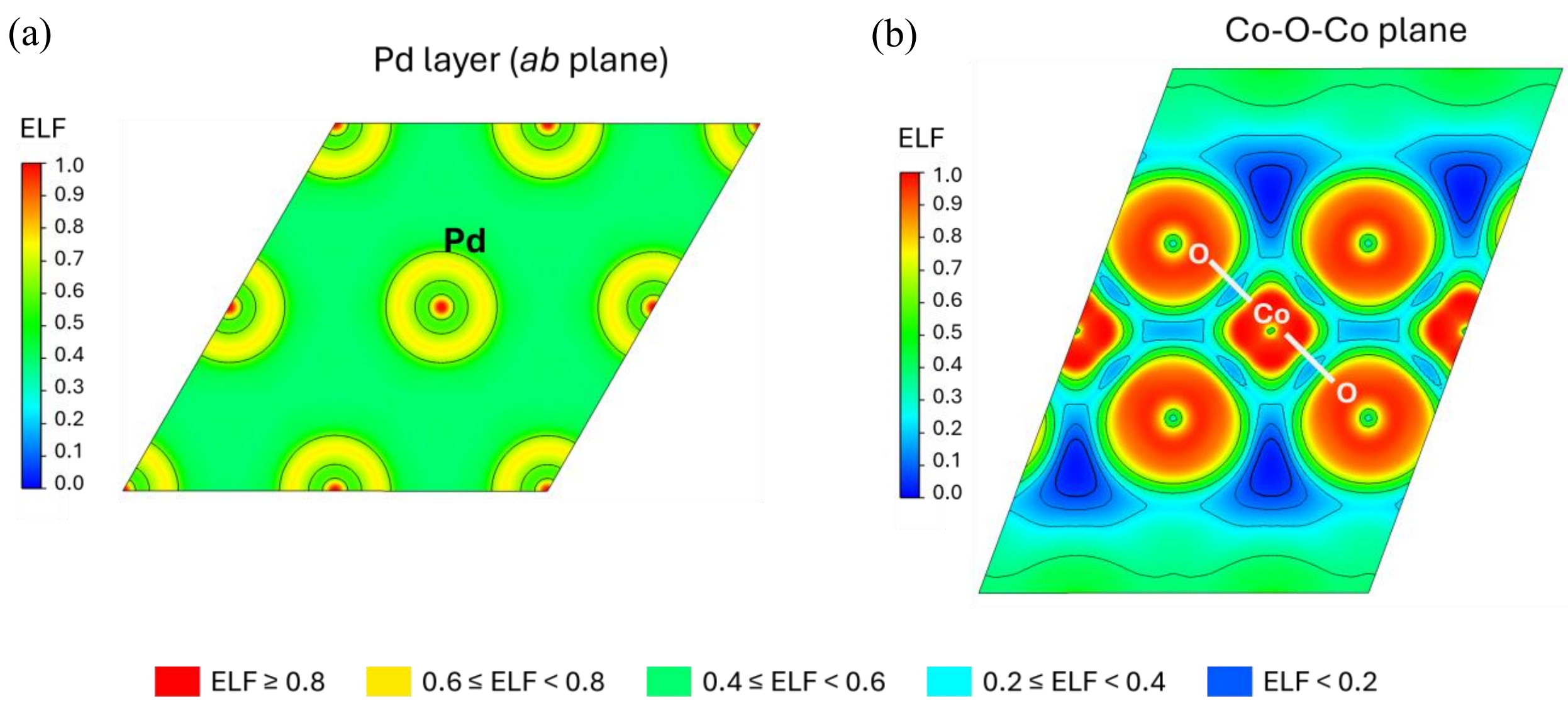

**Figure 3.** Electron localization function (ELF) maps of $PdCoO_2$. ***(a)*** ELF projected onto the Pd layer, showing the delocalized electron distribution characteristic of metallic Pd-Pd bonding. ***(b)*** ELF projected through the $CoO_2$ framework, highlighting the localized electron density associated with strong Co-O bonding. The color scale represents ELF values from 0 to 1.

### F. Electronic robustness of the Pd-derived metallic state

To understand how the anomalous structural response influences the electronic properties, first-principles calculations were performed to examine the pressure evolution of the electronic band structure, projected density of states (PDOS), and Fermi surface. The calculated orbital-projected band structures and PDOS at ambient pressure and 9.9 GPa are shown in **Figure 4**. At ambient pressure, the electronic structure is characteristic of metallic $PdCoO_2$, with a single highly dispersive Pd-derived band crossing the Fermi level ($E_F$). The states near the Fermi level are dominated by Pd 4*d* orbitals, whereas the Co 3*d* and O 2*p* states lie predominantly away from the Fermi level, consistent with the established picture of $PdCoO_2$ as a single-band oxide metal[10,17,35,36]. Upon compression to 9.9 GPa, the overall band dispersion remains nearly unchanged, and the Pd-derived conduction band continues to cross the Fermi level without noticeable band splitting or reconstruction. Likewise, the PDOS exhibits only minor changes in spectral weight with the Fermi-level density of states remaining dominated by Pd orbitals. No additional Co- or O-derived bands emerge at the Fermi level, indicating that pressure does not alter the fundamental electronic character of the conducting channels. A comparison between calculations with and without SOC is provided in the Supplemental Material **Figure S6**. SOC produces only local band splittings and small avoided crossings, without qualitatively changing the density of states near the Fermi level or the Pd-derived metallic band crossing $E_F$, supporting the use of scalar-relativistic calculations for the pressure-dependent analysis. Previous studies have suggested that the Pd-derived conduction band contains Pd 5*s* character [16,37,38]. However, our projected band structure and projected density of states (**Figure S7**) show only a weak Pd 5*s* contribution near the Fermi level. This is attributed to the strong hybridization between the extended Pd 5*s* and Pd 4*d* (particularly $d_{z^2}$) orbitals, resulting in a mixed Pd-derived conduction band rather than distinct 5*s* and 4*d* states.

The calculated Fermi surfaces at ambient pressure and 9.9 GPa (**Figure 5**) further demonstrate the remarkable electronic robustness of $PdCoO_2$. The characteristic quasi-two-dimensional hexagonal Fermi surface is preserved throughout the investigated pressure range, exhibiting only a slight reduction in size associated with lattice contraction. No significant change in Fermi-surface topology or evidence of a pressure-induced Lifshitz transition is observed below 10 GPa. The persistence of the nearly cylindrical Fermi surface confirms that the highly conducting Pd layer remains electronically intact under compression.

(a)

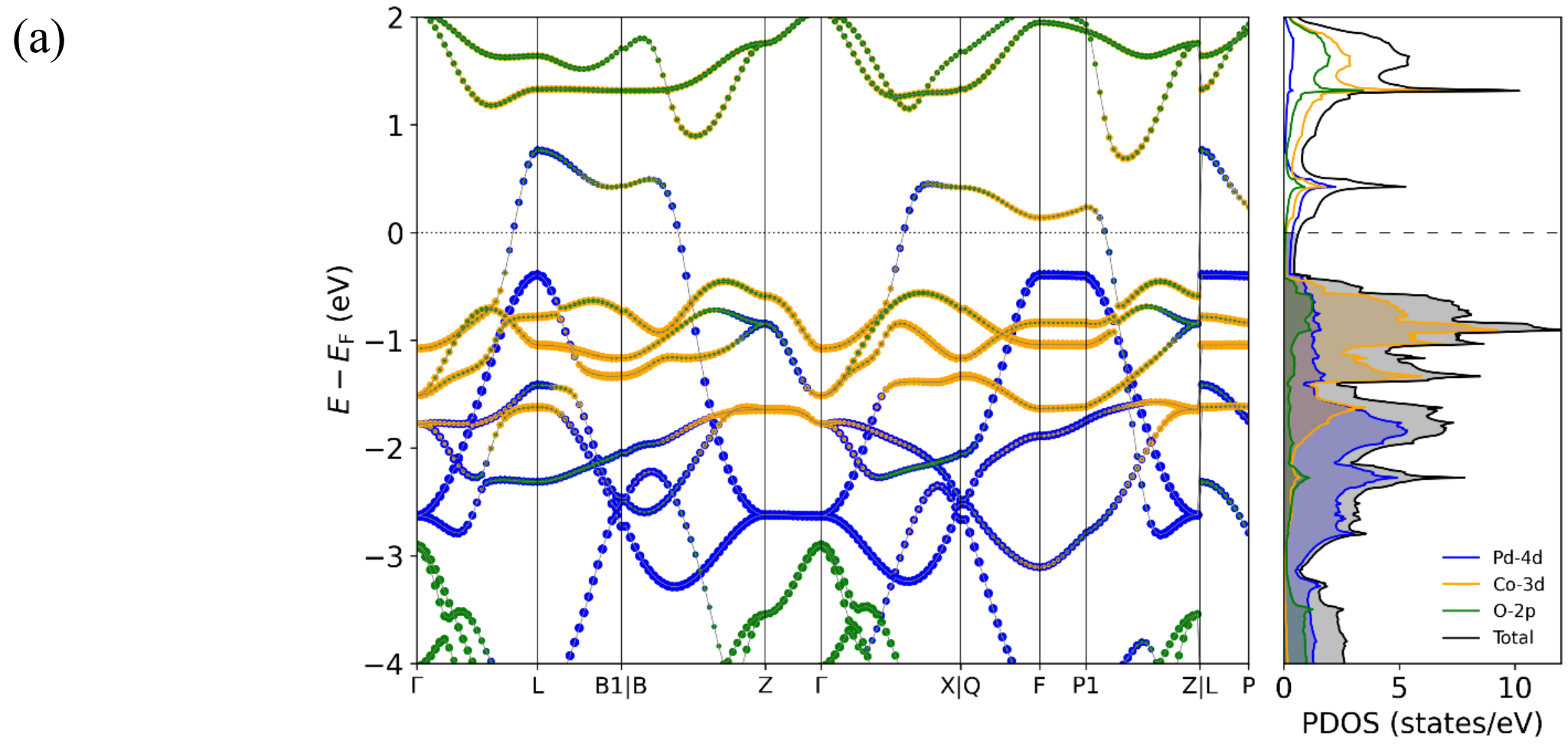

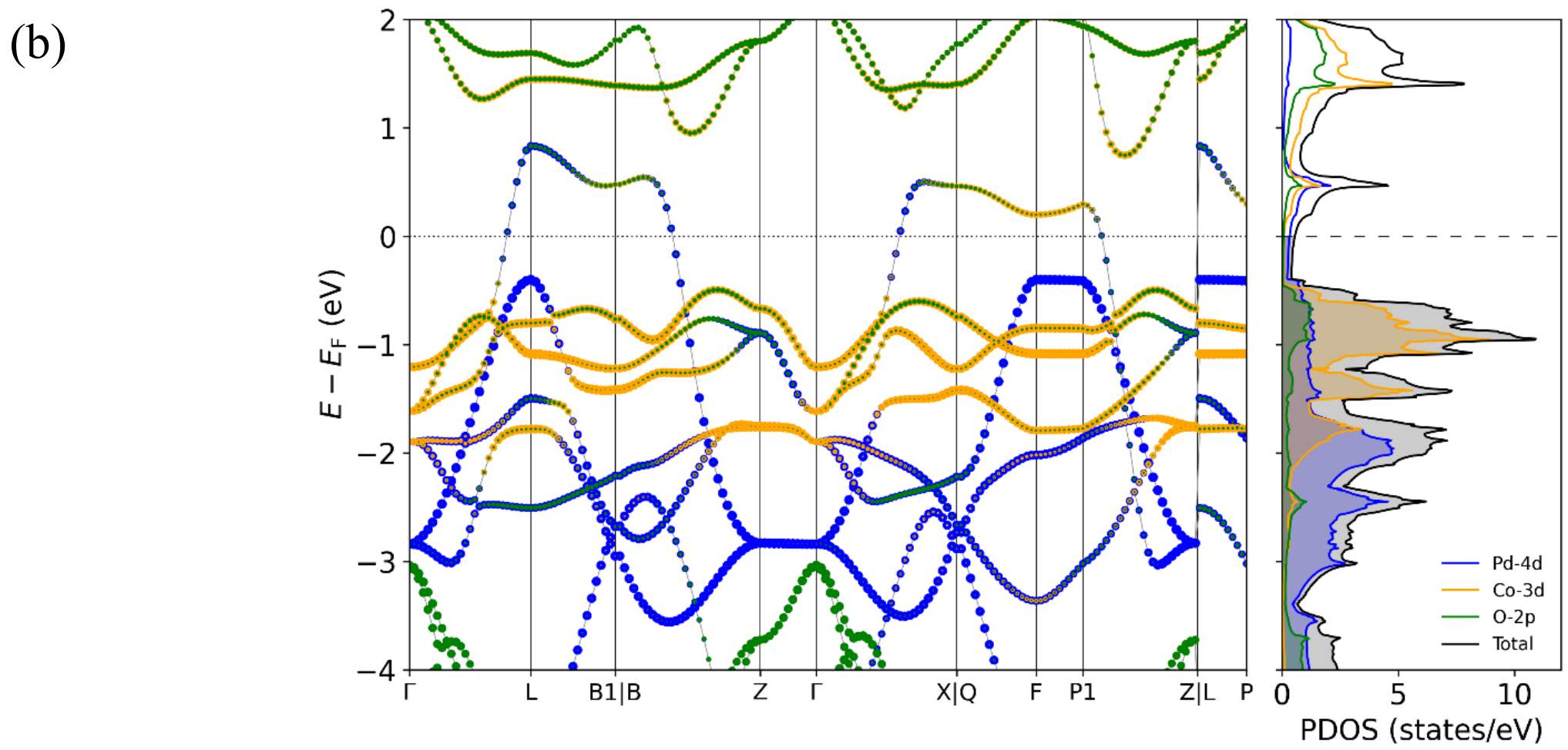

**Figure 4.** Orbital-projected electronic band structures and projected density of states (PDOS) of $PdCoO_2$ calculated at ***(a)*** ambient pressure and ***(b)*** 9.9 GPa.

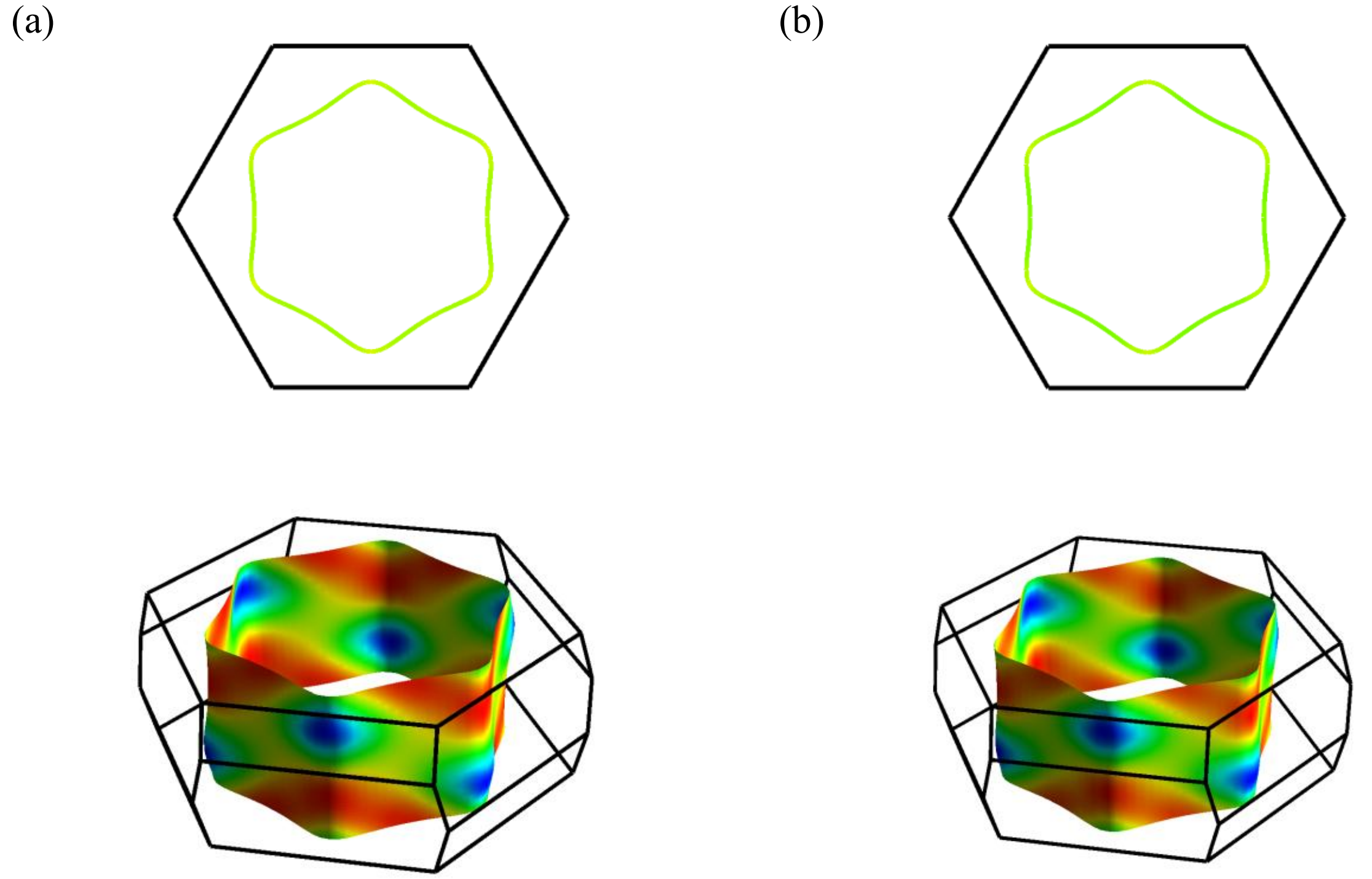

**Figure 5.** Calculated Fermi surfaces of $PdCoO_2$ at ***(a)*** ambient pressure and ***(b)*** 9.9 GPa. The characteristic quasi-two-dimensional hexagonal Fermi surface is preserved under compression, with no apparent change in topology, indicating the absence of a pressure-induced Lifshitz transition below 10 GPa.

## G. Pressure evolution of the transport anisotropy

To evaluate the influence of pressure on the electronic transport properties, Boltzmann transport calculations were performed using BoltzTraP2 within the constant relaxation time approximation. Since the carrier relaxation time ($\tau$) was not explicitly calculated, the results are presented in terms of the normalized conductivity tensor ($\sigma/\tau$) and the conductivity anisotropy rather than absolute conductivity values. In the highly anisotropic quasi-two-dimensional metal $PdCoO_2$, the carrier relaxation time is expected to differ significantly between the in-plane and out-of-plane directions[35]. Consequently, the calculated ratio $[(\sigma_{xx}/\tau + \sigma_{yy}/\tau)/2]/(\sigma_{zz}/\tau)$ represents the anisotropy of the band-structure transport function rather than the true electrical conductivity anisotropy. The experimental conductivity anisotropy, ($\sigma_{ab}/\sigma_c$), additionally depends on the directional relaxation-time ratio, ($\tau_{ab}/\tau_c$), which was not explicitly evaluated in the present work. Therefore, our calculations describe the intrinsic evolution of electronic transport arising from the pressure-dependent band structure, independent of anisotropic scattering effects.

As shown in **Figure 6*a***, the calculated in-plane conductivity, $\sigma_{\text{in-plane}}/\tau = (\sigma_{xx}/\tau + \sigma_{yy}/\tau)/2$, exhibits only a weak pressure dependence throughout the investigated pressure range. At 9.9 GPa, the in-plane conductivity remains approximately 96% of its ambient-pressure value, demonstrating that the Pd-derived metallic conduction channel is largely unaffected by compression. In contrast, the out-of-plane conductivity, ($\sigma_{zz}/\tau$), decreases more substantially, reaching approximately 79% of its ambient-pressure value at 9.9 GPa. As a consequence, the band-structure conductivity anisotropy, ($\sigma_{\text{in-plane}}/\sigma_{zz}$), increases from 3.51 at ambient pressure to 4.26 at 9.9 GPa. Using the experimentally reported ambient-pressure resistivity anisotropy of $\rho_c/\rho_{ab} \approx 150$ at 260 K as a reference[39], this relative increase corresponds to an estimated anisotropy of approximately 180 at 9.9 GPa (**Figure 6*b***), assuming that the relaxation-time anisotropy remains unchanged under pressure. The calculated transport behavior is fully consistent with the structural, bonding, and electronic-structure analyses presented above. The nearly unchanged in-plane conductivity reflects the persistence of the highly dispersive Pd-derived conduction band and the preservation of the quasi-two-dimensional Fermi surface under compression. Meanwhile, the reduction in out-of-plane conductivity indicates that pressure does not promote stronger interlayer electronic transport despite the reduced lattice dimensions. Instead, the enhanced transport anisotropy further confirms that the anomalous compressibility of $PdCoO_2$ originates from its rigid chemical bonding network rather than from an increase in interlayer electronic coupling[6,12,16,39].

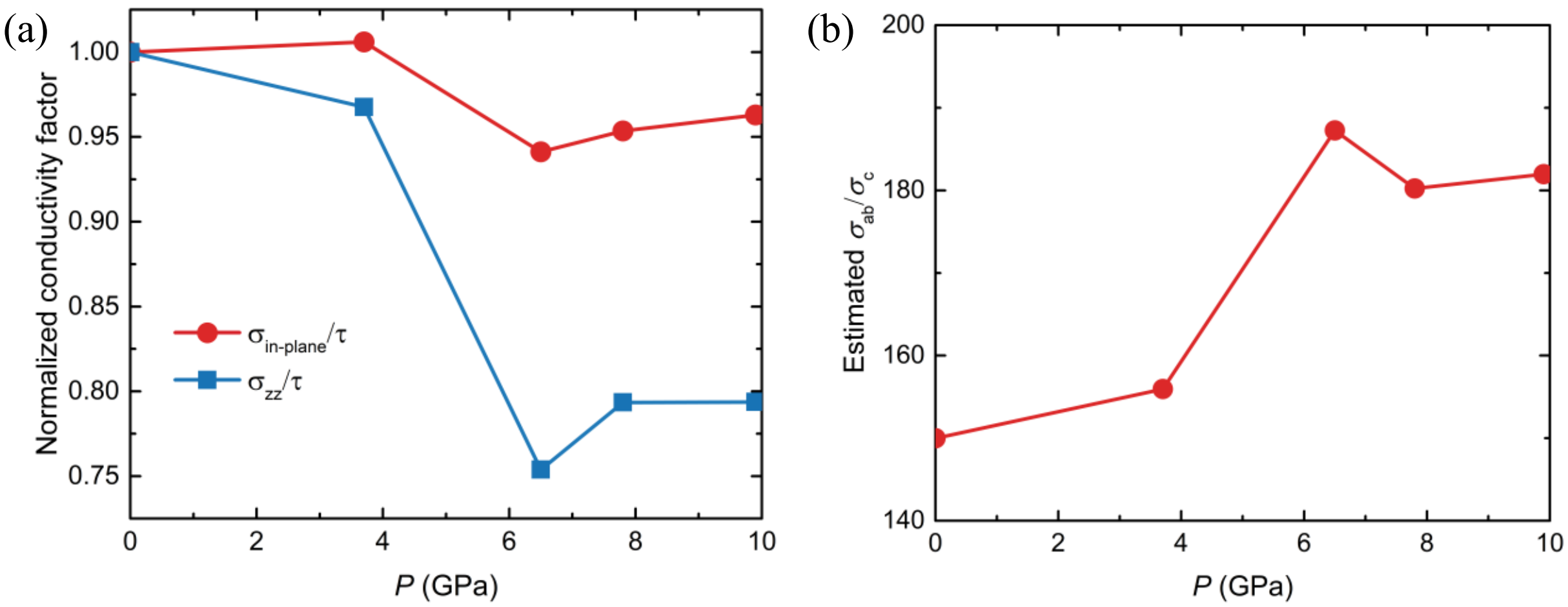

**Figure 6.** Pressure dependence of the calculated transport properties of $PdCoO_2$ obtained from Boltzmann transport calculations. ***(a)*** Normalized in-plane conductivity, ($\sigma_{in\text{-}plane}/\tau$), and out-of-plane conductivity, ($\sigma_{zz}/\tau$), as a function of pressure. ***(b)*** Estimated conductivity anisotropy, ($\sigma_{ab}/\sigma_c$), obtained by scaling the calculated relative change ($\sigma_{in\text{-}plane}/\sigma_{zz}$) to the experimentally reported ambient-pressure value $\rho_c/\rho_{ab} \approx 150$ at 260 K. All calculated values were obtained from BoltzTraP2 at 300 K within the constant relaxation time approximation.

# IV. CONCLUSION

High-pressure single-crystal X-ray diffraction combined with first-principles calculations reveals that $PdCoO_2$ exhibits an unusual pressure response while remaining structurally and electronically robust up to 10 GPa. Despite its layered crystal structure, $PdCoO_2$ displays anomalous anisotropic compressibility, with the in-plane *a*-axis contracting more strongly than the stacking *c*-axis. Structural refinements, COHP, and ELF analyses demonstrate that this behavior originates from a rigid three-dimensional bonding network formed by the linear O-Pd-O units and the $CoO_6$ framework, which preserves the local coordination environment under compression. Correspondingly, the Pd-derived metallic band, quasi-two-dimensional Fermi surface, and highly anisotropic electronic transport remain essentially unchanged, with only a modest increase in transport anisotropy. These results establish $PdCoO_2$ as a pressure-resilient metallic delafossite and provide a microscopic understanding of how robust chemical bonding can govern the mechanical response and stabilize metallic transport in layered oxide conductors.

# ACKNOWLEDGMENTS

The work at Michigan State University was supported by NSF-DMR-2422361. The work at the University of Minnesota was supported by the Department of Energy (DOE) through the University of Minnesota (UMN) Center for Quantum Materials under Grant No. DE-SC0016371.

See **Supplemental Material** for pressure-dependent lattice parameters, bond distances, COHP analysis, electronic structures with the effect of SOC and calculated transport tensor components.

## References

[1] Lynch R W and Drickamer H G 1966 Effect of High Pressure on the Lattice Parameters of Diamond, Graphite, and Hexagonal Boron Nitride *J. Chem. Phys.* **44** 181–4

[2] Aksoy R, Selvi E and Ma Y 2008 X-ray diffraction study of molybdenum diselenide to 35.9 GPa *J. Phys. Chem. Solids* **69** 2138–40

[3] Zhao Z, Zhang H, Yuan H, Wang S, Lin Y, Zeng Q, Xu G, Liu Z, Solanki G K, Patel K D, Cui Y, Hwang H Y and Mao W L 2015 Pressure induced metallization with absence of structural transition in layered molybdenum diselenide *Nat. Commun.* **6** 7312

[4] Shannon R D, Rogers D B and Prewitt C T 1971 Chemistry of noble metal oxides. I. Syntheses and properties of $ABO_2$ delafossite compounds *Inorg. Chem.* **10** 713–8

[5] Shannon R D, Prewitt C T and Rogers D B 1971 Chemistry of noble metal oxides. II. Crystal structures of platinum cobalt dioxide, palladium cobalt dioxide, coppper iron dioxide, and silver iron dioxide *Inorg. Chem.* **10** 719–23

[6] Shannon R D, Rogers D B, Prewitt C T and Gillson J L 1971 Chemistry of noble metal oxides. III. Electrical transport properties and crystal chemistry of $ABO_2$ compounds with the delafossite structure *Inorg. Chem.* **10** 723–7

[7] Ye F, Ren Y, Huang Q, Fernandez-Baca J A, Dai P, Lynn J W and Kimura T 2006 Spontaneous spin-lattice coupling in the geometrically frustrated triangular lattice antiferromagnet $CuFeO_2$ *Phys. Rev. B* **73** 220404

[8] Kawazoe H, Yasukawa M, Hyodo H, Kurita M, Yanagi H and Hosono H 1997 P-type electrical conduction in transparent thin films of $CuAlO_2$ *Nature* **389** 939–42

[9] Mackenzie A P 2017 The properties of ultrapure delafossite metals *Rep. Prog. Phys.* **80** 032501

[10] Eyert V, Frésard R and Maignan A 2008 On the Metallic Conductivity of the Delafossites $PdCoO_2$ and $PtCoO_2$ *Chem. Mater.* **20** 2370–3

[11] Kushwaha P, Sunko V, Moll P J W, Bawden L, Riley J M, Nandi N, Rosner H, Schmidt M P, Arnold F, Hassinger E, Kim T K, Hoesch M, Mackenzie A P and King P D C 2015 Nearly free electrons in a 5d delafossite oxide metal *Sci. Adv.* **1** e1500692

[12] Takatsu H, Yonezawa S, Mouri S, Nakatsuji S, Tanaka K and Maeno Y 2007 Roles of High-Frequency Optical Phonons in the Physical Properties of the Conductive Delafossite $PdCoO_2$ *J. Phys. Soc. Jpn.* **76** 104701

[13] Moll P J W, Kushwaha P, Nandi N, Schmidt B and Mackenzie A P 2016 Evidence for hydrodynamic electron flow in $PdCoO_2$ *Science* **351** 1061–4

[14] Bachmann M D, Sharpe A L, Baker G, Barnard A W, Putzke C, Scaffidi T, Nandi N, McGuinness P H, Zhakina E, Moravec M, Khim S, König M, Goldhaber-Gordon D, Bonn D A, Mackenzie A P and Moll P J W 2022 Directional ballistic transport in the two-dimensional metal $PdCoO_2$ *Nat. Phys.* **18** 819–24

[15] Zhang Y, Saha A, Tutt F, Chaturvedi V, Voigt B, Moore W, Garcia-Barriocanal J, Birol T and Leighton C 2022 Thermal properties of the metallic delafossite $PdCoO_2$: A combined experimental and first-principles study *Phys. Rev. Mater.* **6** 115004

[16] Hicks C W, Gibbs A S, Mackenzie A P, Takatsu H, Maeno Y and Yelland E A 2012 Quantum Oscillations and High Carrier Mobility in the Delafossite $PdCoO_2$ *Phys. Rev. Lett.* **109** 116401

[17] Ong K P, Zhang J, Tse J S and Wu P 2010 Origin of anisotropy and metallic behavior in delafossite $PdCoO_2$ *Phys. Rev. B* **81** 115120

[18] Zhang Y, Tutt F, Evans G N, Sharma P, Haugstad G, Kaiser B, Ramberger J, Bayliff S, Tao Y, Manno M, Garcia-Barriocanal J, Chaturvedi V, Fernandes R M, Birol T, Seyfried W E and Leighton C 2024 Crystal-chemical origins of the ultrahigh conductivity of metallic delafossites *Nat. Commun.* **15** 1399

[19] Garg A B and Rao R 2018 Copper Delafossites under High Pressure—A Brief Review of XRD and Raman Spectroscopic Studies *Crystals* **8** 255

[20] Hasegawa M, Tanaka M, Yagi T, Takei H and Inoue A 2003 Compression behavior of the delafossite-type metallic oxide $PdCoO_2$ below 10 GPa *Solid State Commun.* **128** 303–7

[21] Parkin S, Moezzi B and Hope H 1995 *XABS* 2: an empirical absorption correction program *J. Appl. Crystallogr.* **28** 53–6

[22] Walker N and Stuart D 1983 An empirical method for correcting diffractometer data for absorption effects *Acta Crystallogr. A* **39** 158–66

[23] Sheldrick G M 2015 SHELXT – Integrated space-group and crystal-structure determination *Acta Crystallogr. Sect. Found. Adv.* **71** 3–8

[24] Sheldrick G M 2015 Crystal structure refinement with SHELXL *Acta Crystallogr. Sect. C Struct. Chem.* **71** 3–8

[25] Giannozzi P, Baroni S, Bonini N, Calandra M, Car R, Cavazzoni C, Ceresoli D, Chiarotti G L, Cococcioni M, Dabo I, Dal Corso A, de Gironcoli S, Fabris S, Fratesi G, Gebauer R, Gerstmann U, Gougoussis C, Kokalj A, Lazzeri M, Martin-Samos L, Marzari N, Mauri F, Mazzarello R, Paolini S, Pasquarello A, Paulatto L, Sbraccia C, Scandolo S, Sclauzero G, Seitsonen A P, Smogunov A, Umari P and Wentzcovitch R M 2009 QUANTUM ESPRESSO: a modular and open-source software project for quantum simulations of materials *J. Phys. Condens. Matter Inst. Phys. J.* **21** 395502

[26] Giannozzi P, Andreussi O, Brumme T, Bunau O, Buongiorno Nardelli M, Calandra M, Car R, Cavazzoni C, Ceresoli D, Cococcioni M, Colonna N, Carnimeo I, Dal Corso A, de Gironcoli S, Delugas P, DiStasio R A, Ferretti A, Floris A, Fratesi G, Fugallo G, Gebauer R, Gerstmann U, Giustino F, Gorni T, Jia J, Kawamura M, Ko H-Y, Kokalj A, Küçükbenli E, Lazzeri M, Marsili M, Marzari N, Mauri F, Nguyen N L, Nguyen H-V, Otero-de-la-Roza A, Paulatto L, Poncé S, Rocca D, Sabatini R, Santra B, Schlipf M, Seitsonen A P, Smogunov A, Timrov I, Thonhauser T, Umari P, Vast N, Wu X and Baroni S 2017 Advanced capabilities for materials modelling with Quantum ESPRESSO *J. Phys. Condens. Matter* **29** 465901

[27] Dal Corso A 2014 Pseudopotentials periodic table: From H to Pu *Comput. Mater. Sci.* **95** 337–50

[28] Perdew J P, Burke K and Ernzerhof M 1996 Generalized Gradient Approximation Made Simple *Phys. Rev. Lett.* **77** 3865–8

[29] Monkhorst H J and Pack J D 1976 Special points for Brillouin-zone integrations *Phys. Rev. B* **13** 5188–92

[30] Davidson E R 1975 The iterative calculation of a few of the lowest eigenvalues and corresponding eigenvectors of large real-symmetric matrices *J. Comput. Phys.* **17** 87–94

[31] Setyawan W and Curtarolo S 2010 High-throughput electronic band structure calculations: Challenges and tools *Comput. Mater. Sci.* **49** 299–312

[32] Togo A, Shinohara K and Tanaka I 2024 Spglib: a software library for crystal symmetry search *Sci. Technol. Adv. Mater. Methods* **4** 2384822

[33] Nelson R, Ertural C, George J, Deringer V L, Hautier G and Dronskowski R 2020 LOBSTER: Local orbital projections, atomic charges, and chemical-bonding analysis from projector-augmented-wave-based density-functional theory *J. Comput. Chem.* **41** 1931–40

[34] Madsen G K H, Carrete J and Verstraete M J 2018 BoltzTraP2, a program for interpolating band structures and calculating semi-classical transport coefficients *Comput. Phys. Commun.* **231** 140–5

[35] Yao X, Xun Y, Zhu Z, Zhao S and Li W 2024 Origin of the high electrical conductivity of the delafossite metal $PdCoO_2$ *Phys. Rev. B* **109** 075110

[36] Noh H-J, Jeong J, Jeong J, Cho E-J, Kim S B, Kim K, Min B I and Kim H-D 2009 Anisotropic Electric Conductivity of Delafossite $PdCoO_2$ Studied by Angle-Resolved Photoemission Spectroscopy *Phys. Rev. Lett.* **102** 256404

[37] Tanaka M, Hasegawa M, Higuchi T, Tsukamoto T, Tezuka Y, Shin S and Takei H 1998 Origin of the metallic conductivity in $PdCoO_2$ with delafossite structure *Phys. B Condens. Matter* **245** 157–63

[38] Kim K, Choi H C and Min B I 2009 Fermi surface and surface electronic structure of delafossite $PdCoO_2$ *Phys. Rev. B* **80** 035116

[39] Tanaka M, Hasegawa M and Takei H 1996 Growth and Anisotropic Physical Properties of $PdCoO_2$ Single Crystals *J. Phys. Soc. Jpn.* **65** 3973–7

## Supplementary Information

# Anomalous Compressibility and Electronic Robustness of Metallic Delafossite $PdCoO_2$ under Pressure

Cheng Peng[1], Pahuni Jain[2], Tyler Harper[2], Chris Leighton[2], Weiwei Xie[1*]

3. Department of Chemistry, Michigan State University, East Lansing, MI, 48824 USA
4. Department of Chemical Engineering and Materials Science, University of Minnesota, Minneapolis, MN, 55455 USA

[*] Corresponding Author: Weiwei Xie (xieweiwe@msu.edu)

**Table of Contents**

**Figure S1.** Pressure dependence of the lattice parameters. (**a**) *a* vs. *P*. (**b**) *c* vs. *P*. The SCXRD values (red solid circles), DFT-calculated values (blue open circles) and values from powder data from previous work[20] (black solid squares) are plotted to compare. (**c**) $a/a_0$, $c/c_0$ vs. *P*. (**d**) *c*/*a vs*. P.

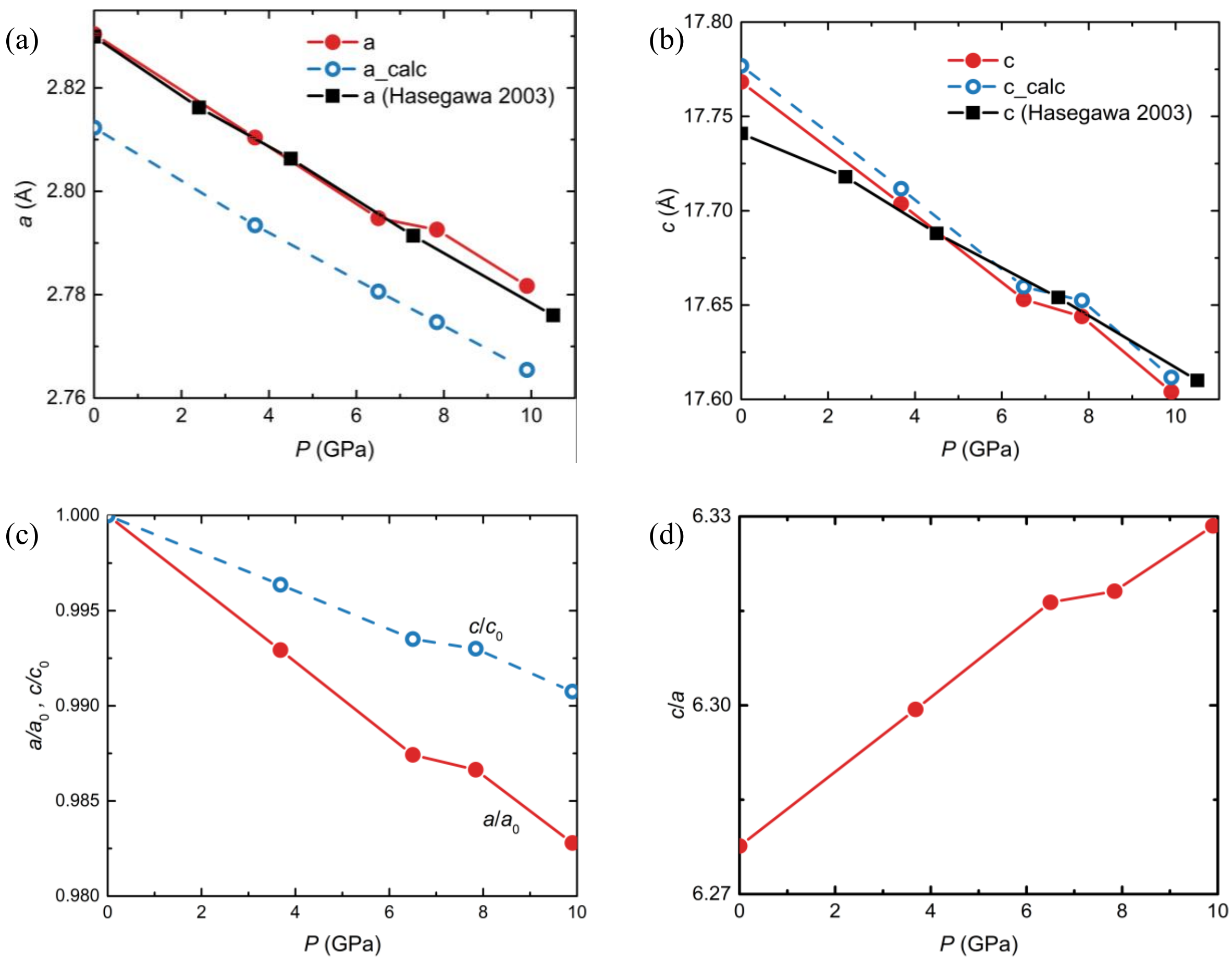

**Table SI.** Lattice parameters *a*, *c*, volume, *V* and ratio, *c*/a of $PdCoO_2$ at each measured pressure at room temperature.

| Pressure (GPa) | *a* (Å) | *c* (Å) | *V* (Å$^3$) | *c*/*a* |
|---|---|---|---|---|
| Ambient pressure | 2.8304(2) | 17.7683(18) | 123.27(2) | 6.27766 |
| 3.7 | 2.8104(2) | 17.7037(12) | 121.096(19) | 6.29935 |
| 6.5 | 2.7948(2) | 17.653(14) | 119.41(2) | 6.31637 |
| 7.8 | 2.7926(2) | 17.644(15) | 119.16(4) | 6.31813 |
| 9.9 | 2.7817(4) | 17.604(2) | 117.97(4) | 6.3285 |

**Figure S2.** Pressure dependence of selected local structural parameters in $PdCoO_2$ determined from high-pressure single-crystal X-ray diffraction: Pd-Pd, Co-Co, Pd-Co, Co-O and Pd-O distances.

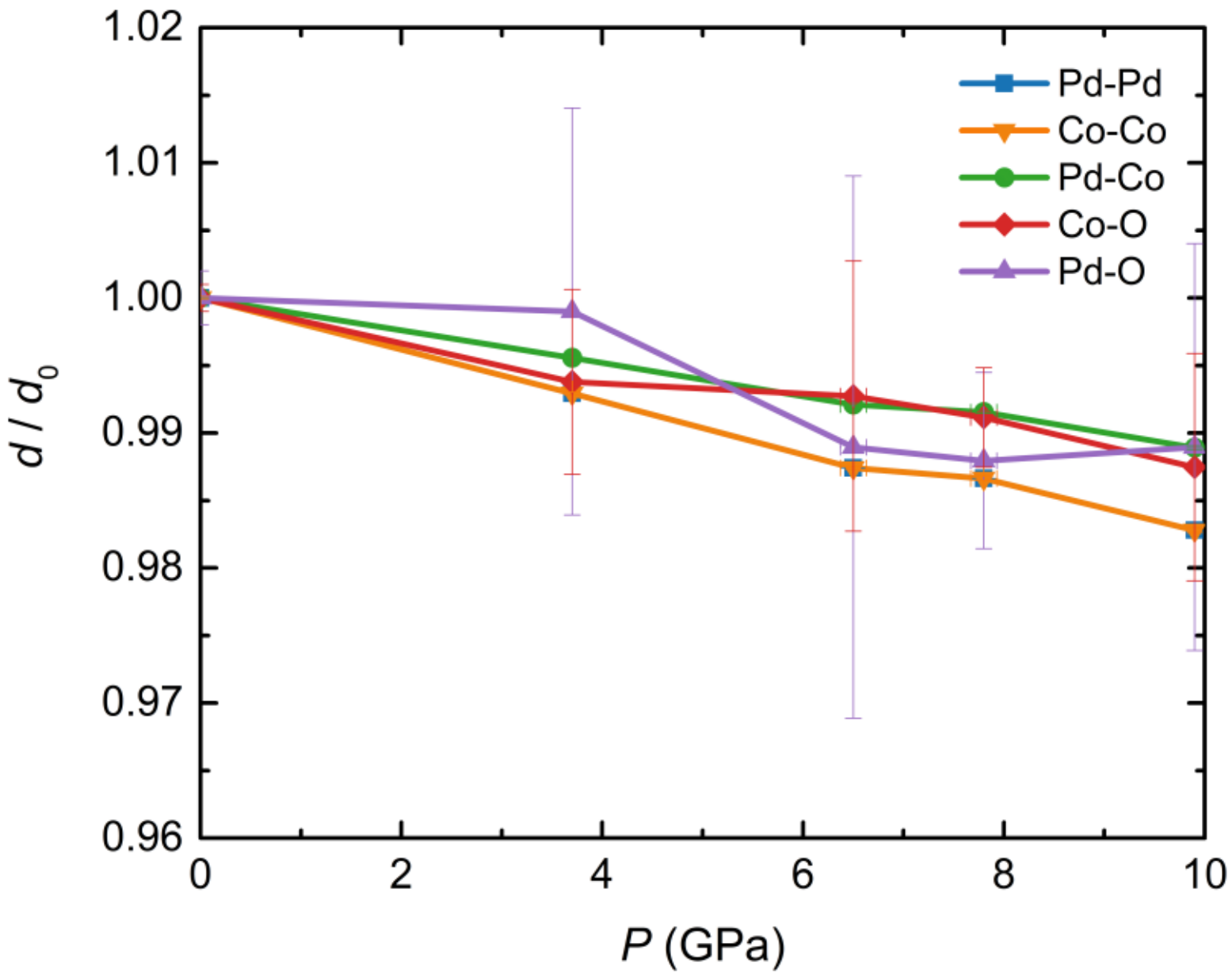

**Table SII.** Bond lengths in $PdCoO_2$ under different pressure.

| Pressure (GPa) | Pd-Pd (Å) | Co-Co (Å) | Pd-Co (Å) | Co-O (Å) | Pd-O (Å) | Co-O-Co (°) |
|---|---|---|---|---|---|---|
| ambient | 2.8304(3) | 2.8304(3) | 3.3823(3) | 1.8998(19) | 1.992(4) | 83.70(13) |
| 3.7 | 2.8104(3) | 2.8104(3) | 3.36733(19) | 1.888(13) | 1.99(3) | 83.8(9) |
| 6.5 | 2.7948(3) | 2.7948(3) | 3.3556(3) | 1.886(19) | 1.97(4) | 84.4(13) |
| 7.8 | 2.7926(4) | 2.7926(4) | 3.3537(3) | 1.883(7) | 1.968(13) | 84.3(5) |
| 9.9 | 2.7817(5) | 2.7817(5) | 3.3448(4) | 1.876(16) | 1.97(3) | 84.3(11) |

**Figure S3.** COHP of Pd-Pd, Co-Co, Pd-Co, Co-O, Pd-O at different pressures.

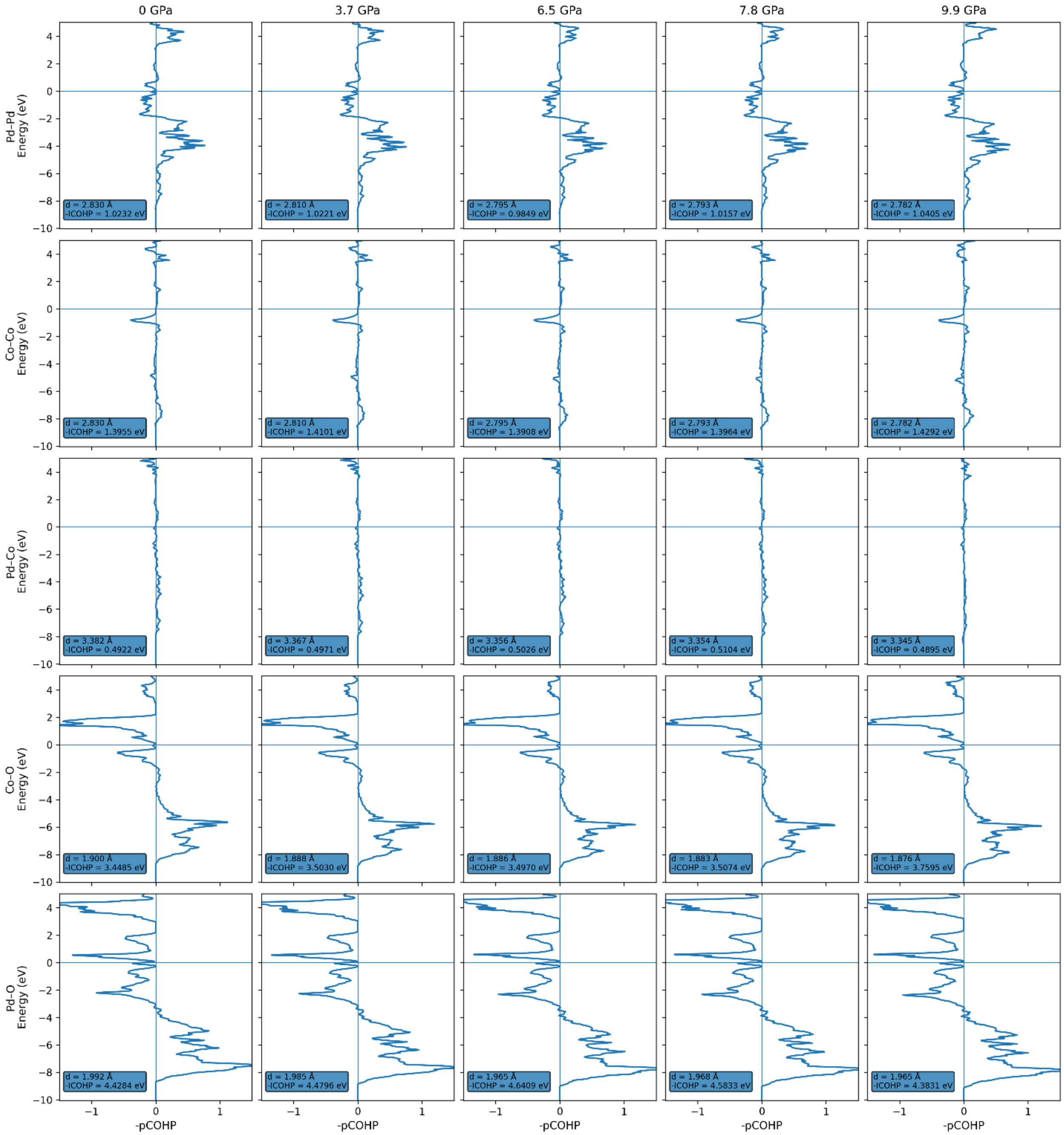

**Table SIII.** The average -ICOHP of the nearest-neighbor pairs under different pressures.

| pair | 0 GPa | 3.7 GPa | 6.5 GPa | 7.8 GPa | 9.9 GPa |
|---|---|---|---|---|---|
| Pd–Pd | 1.0232 | 1.0221 | 0.9849 | 1.0157 | 1.0405 |
| Co–Co | 1.3955 | 1.4101 | 1.3908 | 1.3964 | 1.4292 |
| Pd–Co | 0.4922 | 0.4971 | 0.5026 | 0.5104 | 0.4895 |
| Co–O | 3.4485 | 3.5030 | 3.4970 | 3.5074 | 3.7595 |
| Pd–O | 4.4284 | 4.4796 | 4.6409 | 4.5833 | 4.3831 |

**Figure S4.** Average -ICOHP per bond vs. pressure for (**a**) strong bonds and (**b**) metallic bonds.

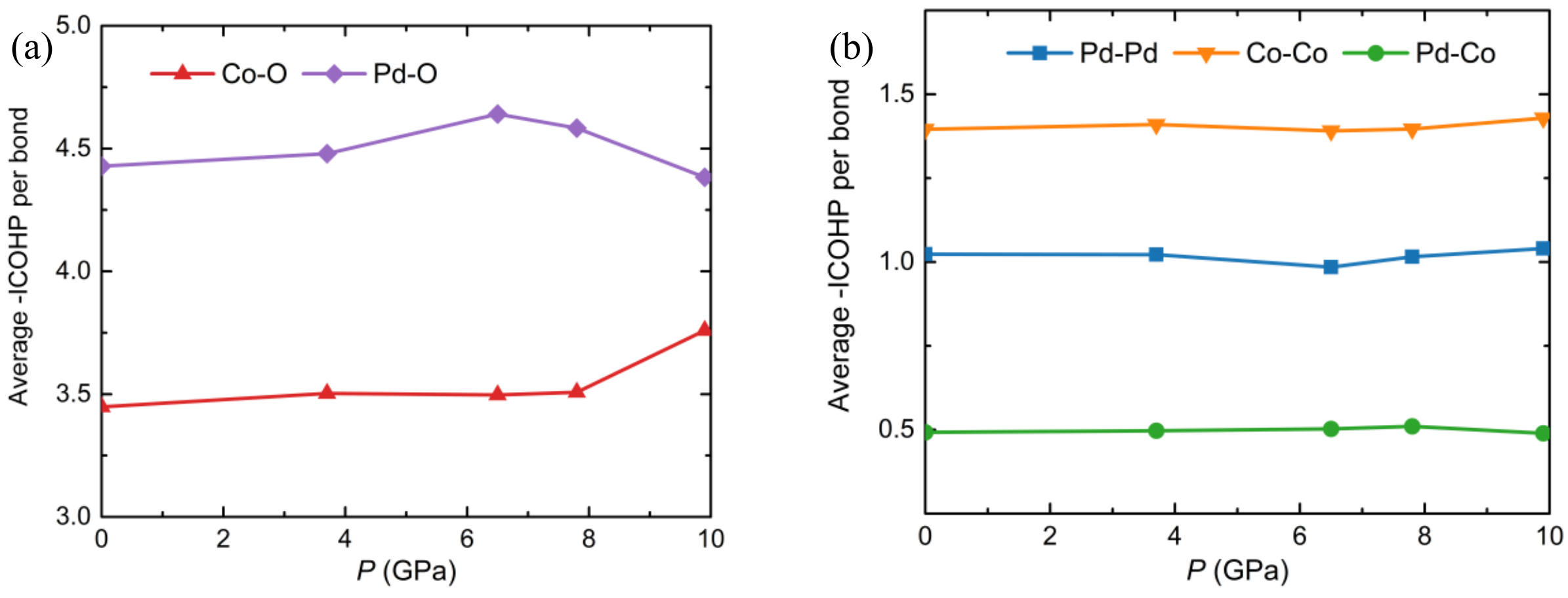

**Figure S5.** Average -ICOHP change at 9.9 GPa relative to 0 GPa.

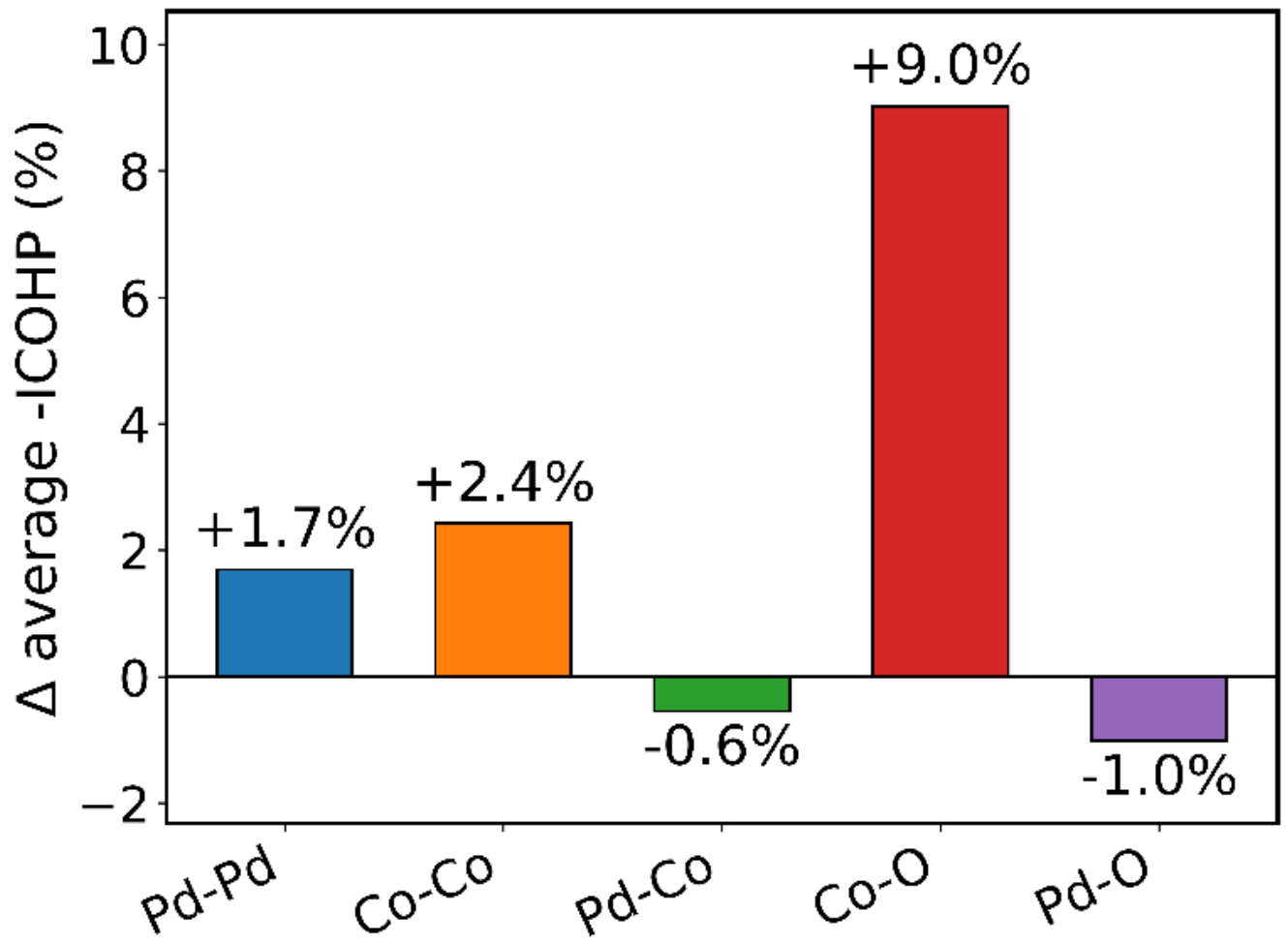

**Figure S6.** Calculated band structures and total density-of-states of $PdCoO_2$ at ambient pressure (**a**) without SOC and (**b**) with SOC.

(a)

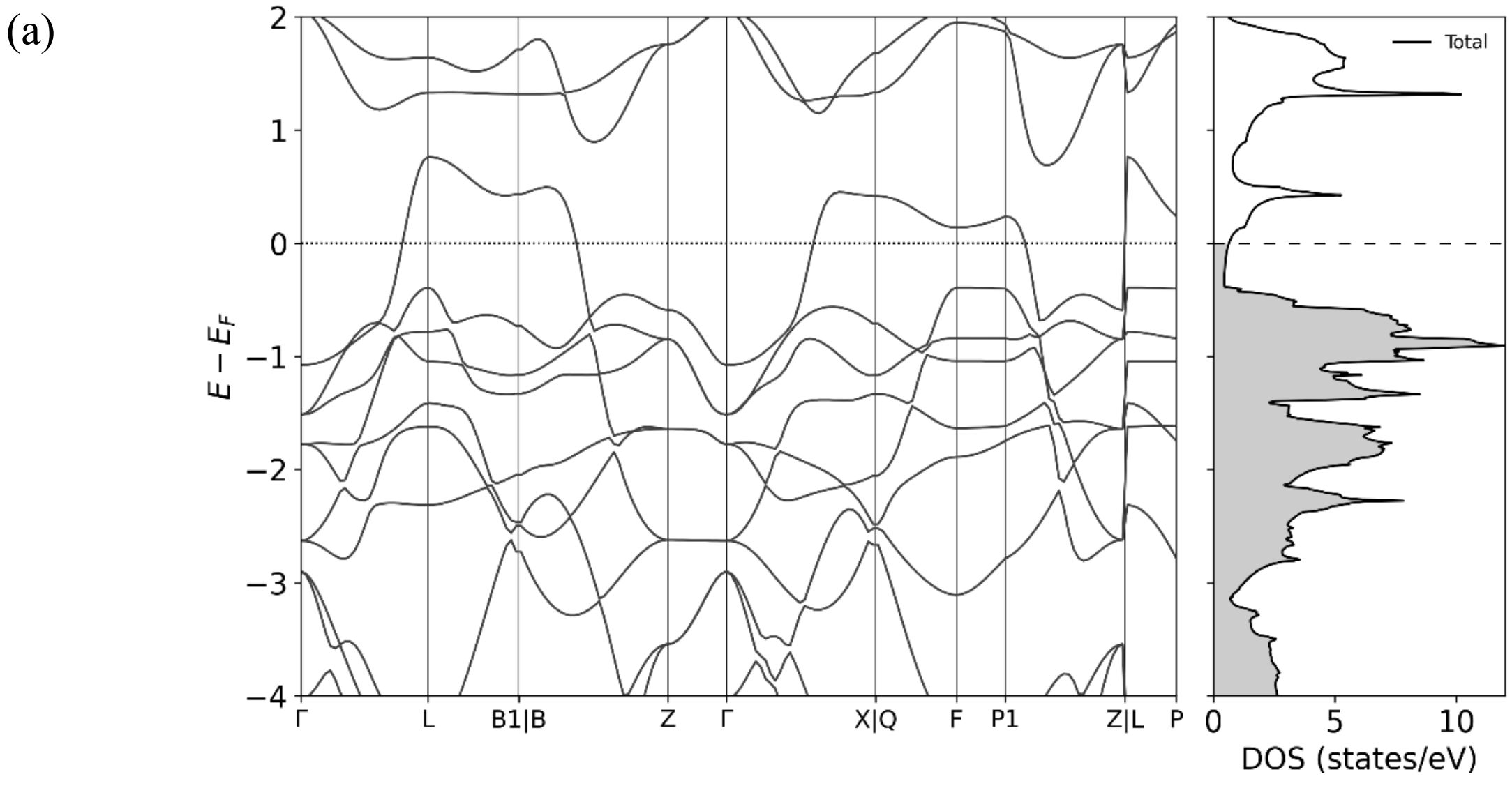

(b)

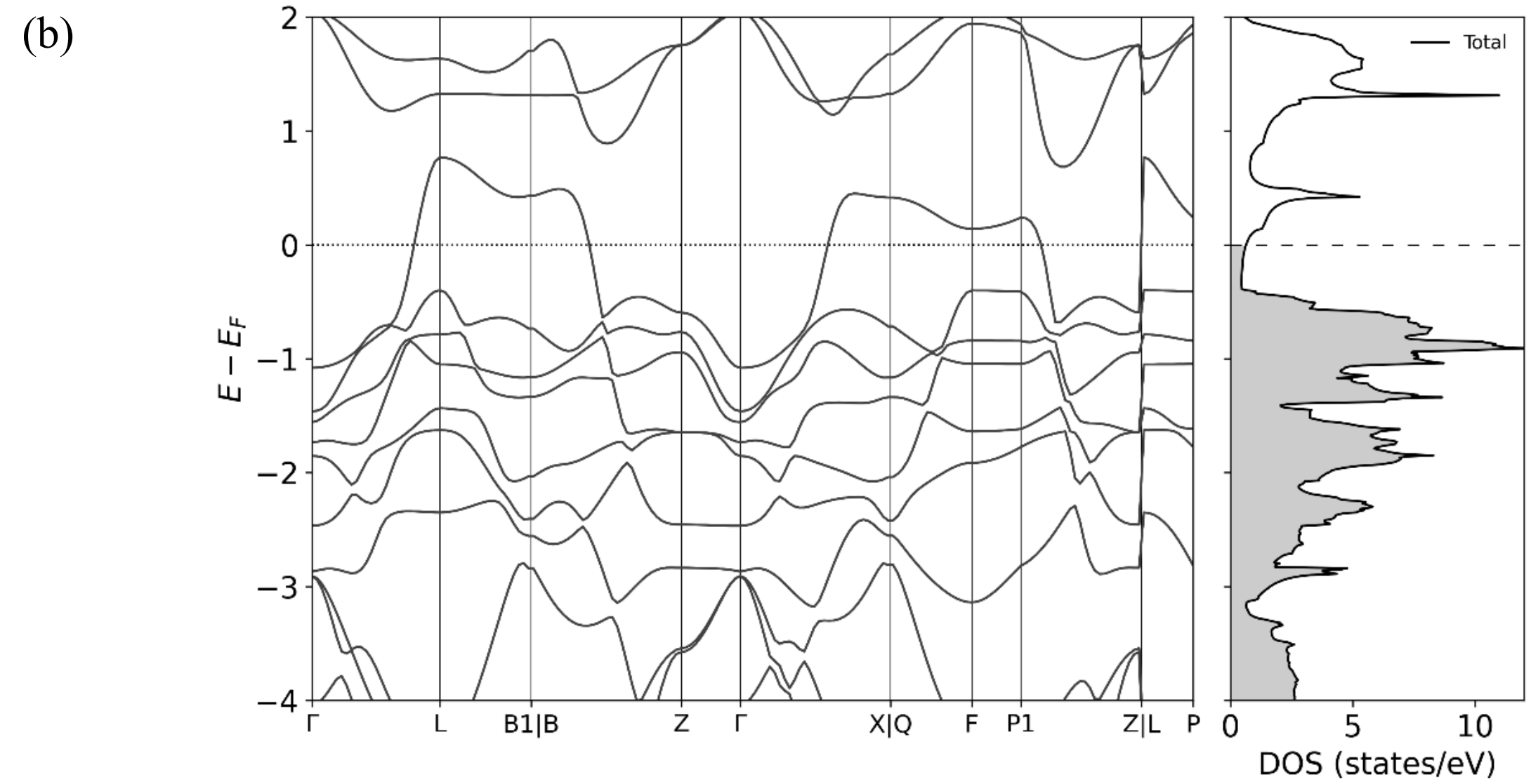

**Figure S7.** Partial Pd band structure and DOS of $PdCoO_2$.

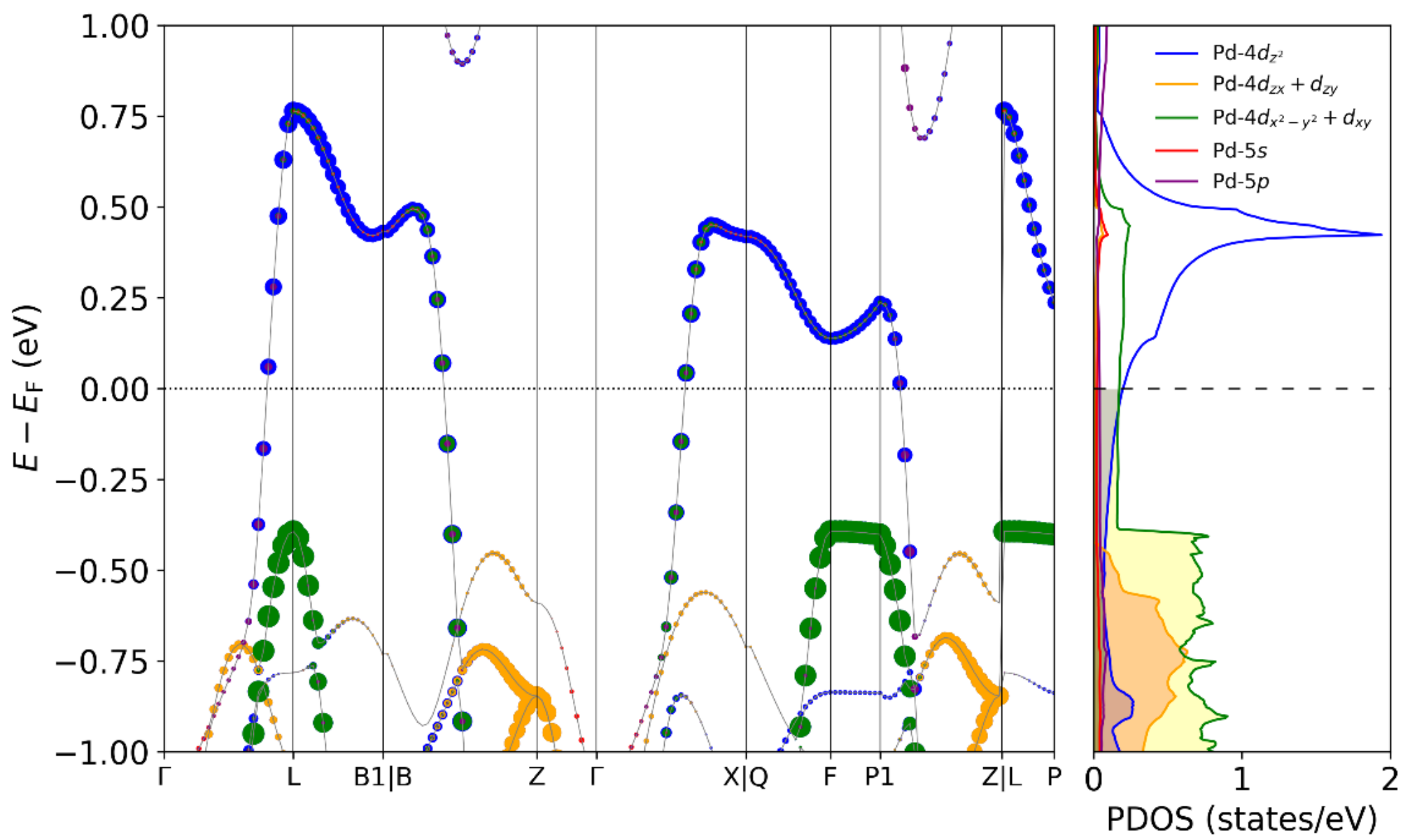

**Table SIV.** Calculated conductivity tensor components for $PdCoO_2$ at different pressures.

| Pressure (GPa) | $\sigma_{\text{in-plane}}/\tau$ ($10^{20}$ $\Omega^{-1}$ $m^{-1}$ $s^{-1}$) | $\sigma_{zz}/\tau$ ($10^{19}$ $\Omega^{-1}$ $m^{-1}$ $s^{-1}$) | $\sigma_{\text{in-plane}}/\sigma_{zz}$ |
|---|---|---|---|
| 0 | 1.08304 | 3.08623 | 3.509 |
| 3.7 | 1.08954 | 2.98618 | 3.649 |
| 6.5 | 1.01933 | 2.32659 | 4.381 |
| 7.8 | 1.03269 | 2.44895 | 4.217 |
| 9.9 | 1.04288 | 2.44967 | 4.257 |